# Implications of Doppler shift for High Frequency Ocean Waves Measured Using Drifting Buoys

W. Erick Rogers,[a] Jim Thomson,[b] and Clarence O. Collins[c]

[a] *Naval Research Laboratory, Stennis Space Center, MS, USA*

[b] *Applied Physics Laboratory, University of Washington, Seattle WA, USA*

[c] *U.S. Army Engineer Research and Development Center, Coastal and Hydraulics Laboratory, Field Research Facility, Duck, NC, USA*

*Corresponding author*: W. Erick Rogers, w.e.rogers.civ@us.navy.mil

ABSTRACT

The advent of expendable wave buoys has greatly expanded the data available for evaluating and calibrating wave models. Ideally, the newer buoys now drifting around the world's oceans would be merged with conventional time series measurements from moored buoys to form a consistent dataset of in situ observations. However, a comparison across several buoy types (moored Datawell, moored NDBC, and two types of drifting buoys) suggests large differences in the high frequency portion of the observed wave energy spectra (0.2 to 0.6 Hz). When binned by wind speed, the moored Datawell buoys have higher energy in the high frequency tail vs. drifting buoys, by factor 1.2 to 1.6. The moored Datawell buoys also have far better agreement with high-frequency energy levels predicted by a numerical wave model. The reference frame of the observations may be the primary contributing factor. To test this hypothesis, the spectra are adjusted from the drifting reference frame to the fixed reference frame. The adjustment is a two-step process, in which the buoy-observed frequencies are first shifted to an intrinsic reference frame, providing wavenumber at each frequency, and then the drifter-observed spectrum is Doppler shifted to the fixed reference frame. Two methods for estimation of buoy drift are tested; one is based on wind speed, and one is based on buoy positions. With this adjustment, the observations from the drifting buoys become more consistent with the moored Datawell buoys, though discrepancies still exist with the moored NDBC buoys.

SIGNIFICANCE STATEMENT

The aim of this study is to resolve an apparent discrepancy between ocean wave spectra measured drifting buoys vs. those measured using moored buoys. Since ocean wave direction has significant correlation with buoy drift direction, one possible explanation is that the discrepancy, which occurs primarily at higher frequencies, can be attributed to Doppler shift. Results of our analysis support this hypothesis. We also demonstrate that correction for this effect is possible. Such correction is useful and important, since these observations are routinely used to calibrate and correct systems that predict ocean waves. Such predictions have multiple, overlapping civilian and military applications.

# 1. Introduction

Observations of ocean waves by buoys and satellites have long been the primary "ground truth" by which wave prediction models are evaluated and calibrated. They are also used to improve prediction systems in real time through data assimilation (e.g., Voorips et al., 1997; Houghton et al., 2022). Furthermore, these observations are a key component in the construction of parametric models (e.g., Battjes et al., 1987), climatologies (e.g., Liu et al., 2023), and new machine-learning-based predictions (e.g., Bodnar et al., 2025), either by direct inclusion or indirectly via the numerical models used to create these products.

However, inconsistencies between buoy datasets (e.g., Jensen et al., 2021) have long confounded their usage, particularly when an application is sensitive to small discrepancies (e.g., Liu et al., 2023) or when focusing on specific features of wave spectra (e.g., O'Reilly et al., 1996). These inconsistencies can be introduced by differences in platforms, instrumentation, and mooring designs. In the case of model calibration, for instance, such inconsistencies imply that calibrating to one dataset will necessarily degrade agreement with another; analogous problems are introduced for the other applications listed above.

In this study, we focus on the high-frequency region of the gravity wave spectrum, roughly 0.2 to 0.6 Hz. Although there are some exceptions (e.g., Ardhuin et al., 2010; Wang et al., 2020), these shorter waves are primarily observed using buoys rather than satellites. Over the last decade, an increasingly large fraction of the worldwide ocean wave buoy population has shifted toward drifting buoys, such as the extensive network of Spotters manufactured and managed by Sofar Ocean Technologies (Houghton et al., 2021). As will be detailed in Section 2, these new observation types record high-frequency energy levels that are inconsistent with at least one major moored buoy network. In this paper, we explore the hypothesis that this discrepancy can be explained as an outcome of Doppler shift, since drifting buoys have a markedly different frame of reference compared to traditional moored buoys. Because Doppler shift is roughly proportional to the square of the frequency ($f^2$), it is especially relevant to the high-frequency waves targeted in this study.

The impact of Doppler shift on buoy observations is not a new topic. Even for moored buoys, it can be an important consideration due to local currents (Forristall et al., 1978; Battjes et al., 1987). Researchers have also studied the impact of Doppler shift on propelled platforms, such as ships (e.g., Cartwright, 1963; Drennan et al., 1994; Hanson et al., 1997; Cifuentes-Lorensen et al., 2013; Collins et al., 2017; Olberg et al., 2024), Autonomous Surface Vehicles (ASVs; Amador et al., 2023; Colosi et al., 2023) and scanning radar (Walsh et al. 1985). Because of the larger speeds of these platforms, the corresponding Doppler effects are stronger. Although drifting buoys travel at lower speeds—and thus experience a smaller Doppler shift—the sheer volume of ocean wave data generated by these buoys vastly exceeds that of ships and ASVs. Yet, the impact on drifting buoys has received little attention. We note two recent exceptions: Davis et al. (2025) discussed and quantified the problem during a study of mean square slope observed under a hurricane, and Mounet et al. (2026) proposed a generalized tool for estimating Doppler shift applicable to any moving platform.

To address this new problem, we incorporate three noteworthy features into our methodology. First, we account for three frames of reference (intrinsic, fixed, and drifting) rather than the standard two. Second, we employ an exceptionally large dataset comprising more than three million spectra from Spotter buoys alone. Third, we quantify—in a mean sense—how this Doppler-shift effect varies with wind speed, allowing us to compare different buoy types.

The paper is organized as follows. In Section 2, we present the inconsistency in high frequency energy found with different buoy types and offer two hypotheses to explain the discrepancy. In Section 3, the methods used for Doppler-shift correction with three frames of reference are described. Results and discussion are presented in Sections 4 and 5, respectively. In Section 6, the main findings are summarized.

## 2. Background

### *a. Initial comparisons and apparent discrepancies*

While performing validation of an ocean wave model, WAVEWATCH III® (“WW3”, Tolman 1991; WW3DG 2019) against buoy data, we noticed a peculiar trend: apparent bias in high frequency energy level when validating against drifting buoys indicated a large positive bias, while validation against a moored buoy indicated a small negative bias (for details, see Rogers and Thomson 2026). In other words, the primary conclusions from the validation were confused and contradictory, and no decision was made for follow-up calibration. Our main purpose here is to pull on these threads and resolve this inconsistency.

The process described above, comparing {model vs. buoy} vs. {model vs. buoy} is valid but indirect method of comparing two buoy datasets. Similarly, direct time-series comparison of co-located buoys is impractical: a drifting buoy will not stay co-located with a moored buoy for long. Fortunately, since we are interested in high frequency ocean waves, a shortcut is available to us. In general, a high frequency wave component is more likely to be fully developed than a low frequency wave component[1]. Such components are not limited by fetch, ($\partial E/\partial x \approx 0$), nor limited by duration ($\partial E/\partial t \approx 0$) and are thus in equilibrium state, with net source terms in balance with the local winds ($\sum S = S_{in} + S_{ds} + S_{nl} \approx 0$).[2] Thus, the energy level $E$ is purely a function of wind speed, $U_{10}$. This is, of course, approximate. First, the relations just given merely *approach zero* at high frequencies. Second, environmental features may affect the source terms balance, such as bathymetry, horizontally sheared currents, and atmospheric stability. Further, the energy flux from wind to waves is determined by the wave-supported stress, which is not a simple function of 10-m wind speed. Such factors result in scatter for the $E$ vs. $U_{10}$ relation.

In this study, we use two methods to define the wave component. One is to use the non-directional spectral density at a test frequency, $E(f)$. Another is to compute the energy in a frequency band, and then convert to an equivalent wave height, i.e., the "band height": $H_{m0B} = 4\sqrt{m_{0B}}$ where $m_{0B} = \int_{f_1}^{f_2} E(f)df$. An example of $H_{m0B}$ plotted against wind speed is shown in Fig. 1, using $f_1 = 0.29$ Hz and $f_2 = 0.58$ Hz. The observational dataset is from drifting Spotter buoys deployed by Sofar Ocean. This large buoy network measures wave spectra over much of the world's oceans. It was provided by P. Smit of Sofar for the period September 2021 to May 2022. This is a very large dataset: over three million spectra are represented in Fig. 1 ($n = 3.27 \times 10^6$). The 10-m neutral wind speeds are taken from the ECMWF (European Centre for Medium-Range Weather Forecast) Reanalysis v5 (ERA5) (Hersbach et al. 2020). Fields are provided at 1-hourly intervals and 1/4° geographic resolution. We use nearest neighbor in time and bilinear interpolation in space for co-location.

As is visible in Fig. 1, the data are binned in wind speed bins of width 1 m/s. The black line shows the mean $H_{m0B}$ in each wind speed bin.

Fig. 1 illustrates the degree of scatter about the mean trend line. In subsequent plots, we use mean trend lines to visualize and compare multiple datasets.

---

[1] For example, Chen et al. (2016) study the response time of a metric that is strongly dependent on high frequencies, mean square slope, $mss$, which, in deep water, is proportional to the fourth moment of the wave spectrum: $mss \propto m_4 = \int E(f) f^4 df$. They find that $mss$ has a typical response time of only one hour to changes in wind speed.

[2] Here, $x$ is fetch, $S_{in}$ indicates positive source terms (energy flux from winds to waves), $S_{ds}$ indicates negative source terms, such as dissipation by whitecapping, and $S_{nl}$ indicates energy-conserving source terms, such as four-wave nonlinear interactions.

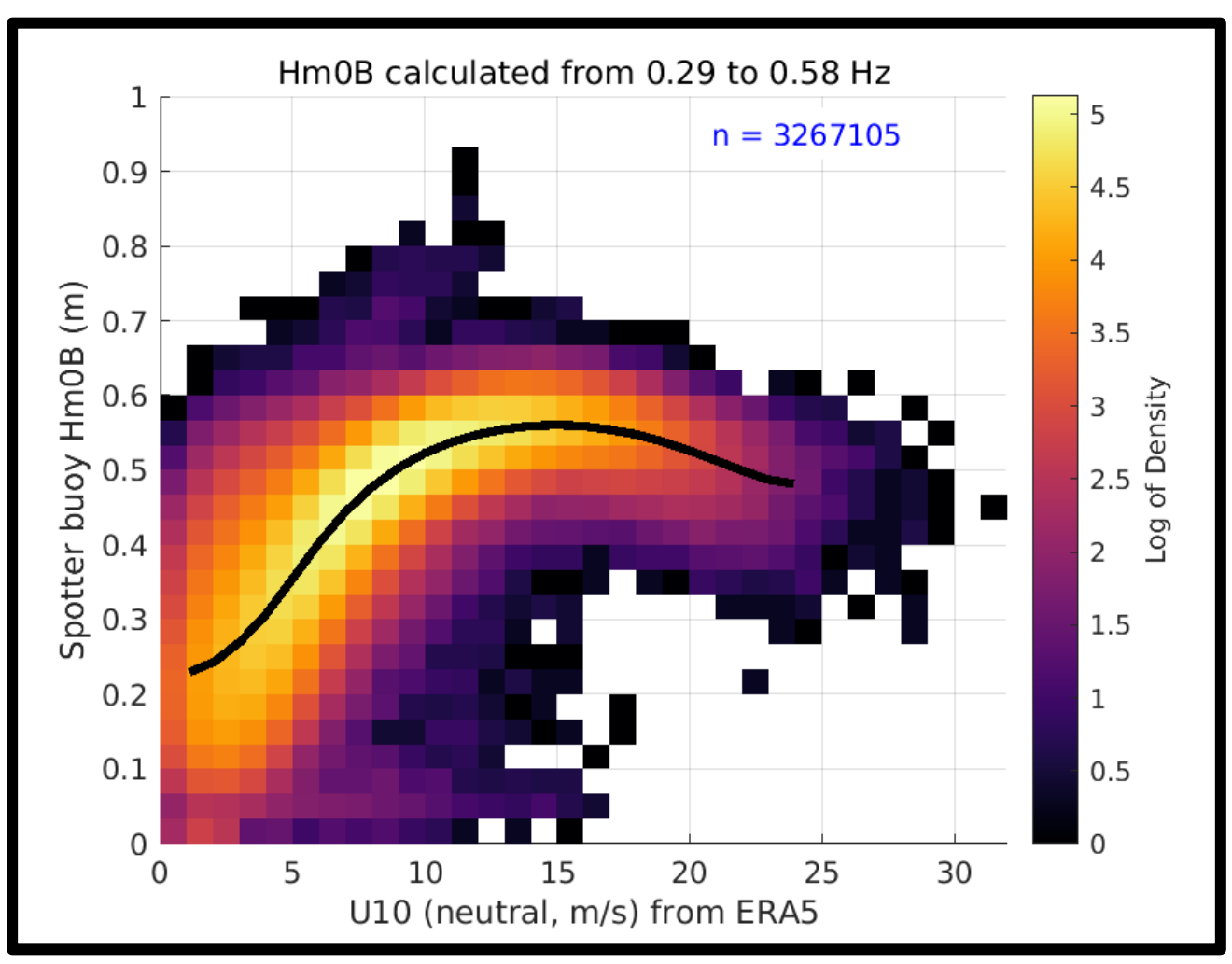

Fig. 1. High-frequency band height, $H_{m0B}$, as a function of wind speed, based on nine months of observations from Sofar drifting buoys. Color scale indicates the number of occurrences in each bin (2-d histogram, log10 scale). The black line indicates the mean for each wind speed bin. [Reproduced with minor changes from Rogers and Thomson (2026).]

Fig. 2 compares the mean trend lines for $H_{m0B}$ in for the following datasets:

1) A smaller ($n = 1.30 \times 10^4$) drifting "mini-wave-buoy" dataset deployed by University of California, San Diego (UCSD) Scripps Institute of Oceanography (SIO) Coastal Observing Research and Development Center (CORDC), provided by E. Terrill, S. Merrifield, and others. The time period of the data is 24 November 2016 to 11 April 2017 (thus, 4.6 months duration). All spectra were measured in the Pacific Ocean, between 25°N and 35°N.
2) A co-location of a WW3 hindcast with (1). The model is forced by ERA5 10-meter neutral winds and has approximately 1/4° geographic resolution. Currents are not included. Details of the model setup are provided in Rogers and Thomson (2026).
3) Data from a Datawell Waverider (DWR) buoy deployed in the north Pacific (50°N, 145°W) at Ocean Station Papa (OSP), operated by the U. Washington Applied Physics Laboratory (UW/APL) and published by the Coastal Data Information Program (CDIP), which is affiliated with UCSD. The time period is a full year, 16 May 2022 – 17 May 2023, $n = 1.72 \times 10^4$.
4) A co-location of a WW3 hindcast with (3). The setup of the model is the same as that of (2).
5) The Sofar dataset shown in Fig. 1, $n = 3.27 \times 10^6$.
6) A one-month subset of (5), February 2022, $n = 3.41 \times 10^5$.
7) A co-location of a WW3 hindcast with (6). The setup of the model is the same as that of (2).

Fig. 2 illustrates the striking discrepancy described at the beginning of this section: WW3 results are consistent with the OSP DWR buoy, while the drifting buoy datasets have much lower energy levels.

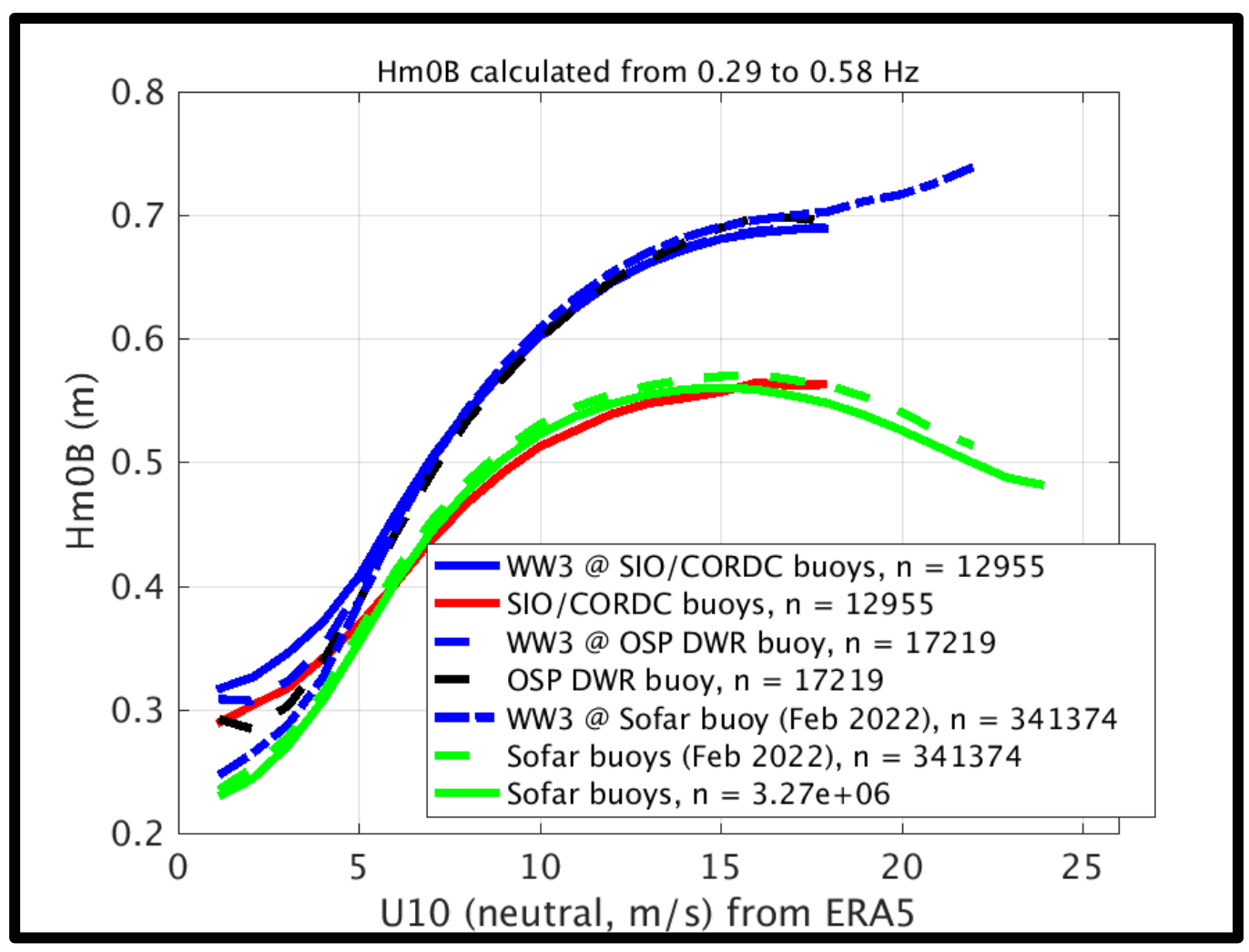

Fig. 2. Comparison of the mean trend lines of high-frequency band height, $H_{m0B}$ vs. wind speed. Blue lines are from WW3 hindcasts, co-located with buoy datasets. Black, green, and red lines are from a Datawell buoy, Sofar buoys, and CORDC buoys, respectively. Reproduced with minor changes from Rogers and Thomson (2026).

In Fig. 3, $E(f)$ at $f$ =0.4 Hz is evaluated, replacing $H_{m0B}$ used in Fig. 2. Datasets (6,7) are omitted, and two new datasets are introduced:

8) Data from 13 UCSD/SIO/CDIP buoys including the OSP buoy, for up to 32 months duration per buoy, depending on outages, May 2022 to December 2024. Data were downloaded from the CDIP website. Locations other than OSP are near US coasts, including Hawaii and American Samoa.

9) Data from eight National Oceanic and Atmospheric Administration (NOAA) National Data Buoy Center (NDBC) buoys, for up to 12 months duration per buoy, depending on outages, January to December 2024. Data were downloaded from the NDBC website. Locations are: three near the US east coast, two near the US south coast, two near Alaska, and one near Hawaii.

For plots of the geographic distributions of these nine datasets, we refer the reader to Figures 1 and 2 of Rogers and Thomson (2026).

With the exception of the NDBC dataset (9), behavior of drifting and moored buoys in Fig. 3 is broadly similar to that of Fig. 2. The NDBC dataset (9) has systematically lower $E(f)$ than the moored DWR datasets (3,8), Further, this energy level is lower than that of the drifting buoys for $U_{10} < 14$ m/s. There is agreement between moored NDBC and the drifting buoys for $U_{10}$ =14 to 18 m/s. This is superficially encouraging, but in Section 4, we will present evidence that this limited region of intersection is produced by a fortuitous cancellation of errors.

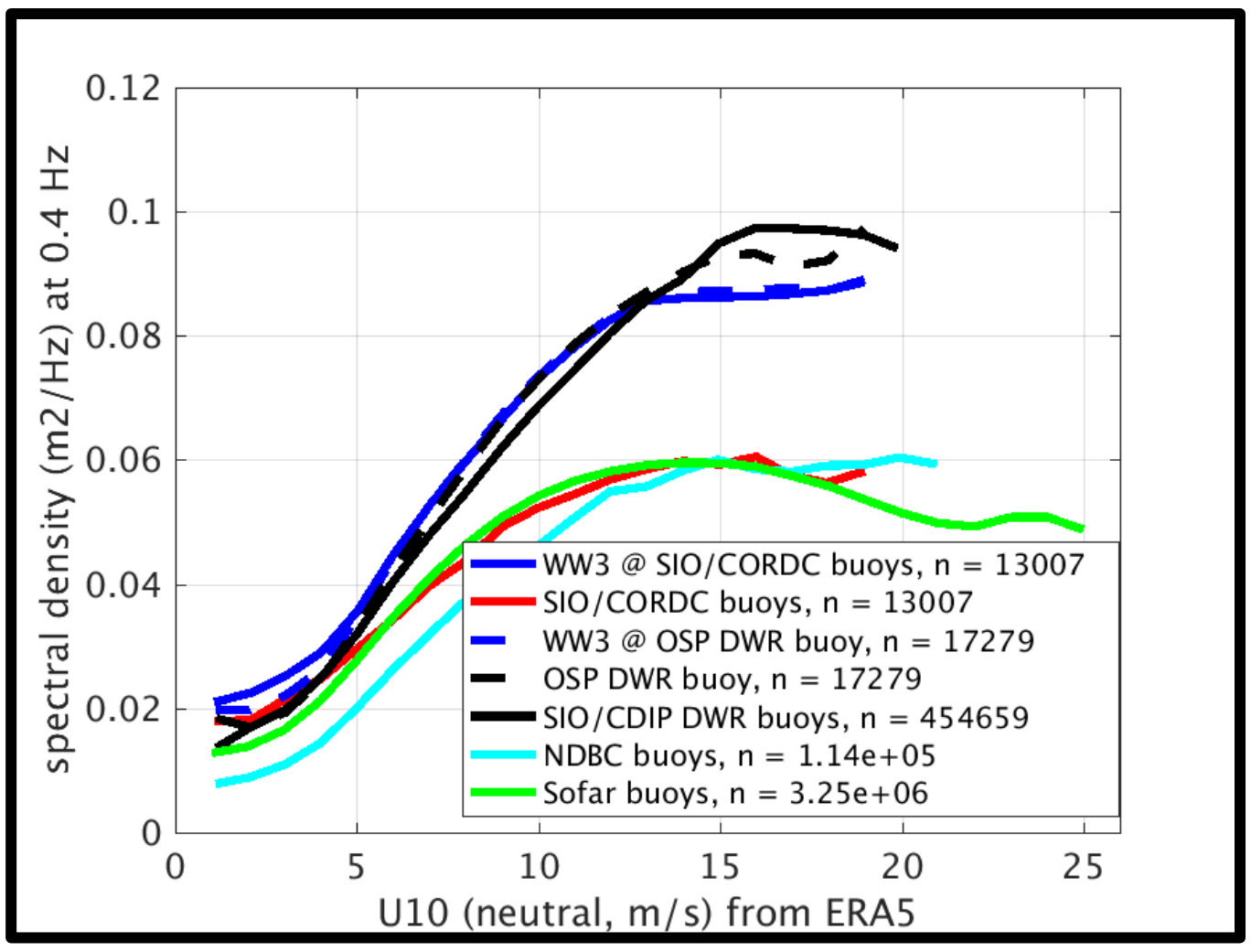

Fig. 3. Like, Fig. 2, but showing the mean value lines of $E(f)$ at $f$=0.4 Hz. Modified from a figure in Rogers and Thomson (2026).

### *b. Hypotheses, to explain discrepancies*

We propose two possible explanations for the discrepancies summarized in Fig. 2 and Fig. 3.

*Hypothesis 1: Discrepancy is caused by differences in buoy hardware and software.* This includes effects of buoy/hull response, missing or incorrect application of corrections for buoy response, instrumentation, and signal processing methods. Buoy hydrodynamic response is quantified using a response amplitude operator (RAO), where deviation of the RAO from unity indicates that correction is needed. For example, Ribe (1982) presents an RAO of an early DWR buoy and found that it is near unity until around 0.5 Hz, beyond which there is an increasing resonant excitation (over-response). Large buoys, such as 3-m foam and discus buoys of NDBC, can be expected to have a damped response for short waves; NDBC does not report $E(f)$ after $f$ =0.485 Hz for these buoys. Mooring design and biofouling may also affect buoy response at high frequencies (Thomson et al. 2015). We do not know of specific reasons to attribute discrepancies to differences in instrumentation (e.g., using GPS vs. Inertial Measurement Unit (IMU)) or processing methods, e.g., noise filters, in the context of high frequency waves.

The current study does not explore this hypothesis. For further reading, see Ribe (1982), Thomson et al. (2015), Jensen et al. (2021), Collins and Jensen (2022), Collins et al. (2024), Rogers and Thomson (2026).

*Hypothesis 2. Discrepancy is caused by Doppler shift.* In the open ocean, a scenario where buoy drift is aligned with high-frequency wave direction is common since both are strongly affected by wind. In this scenario, waves observed by drifting buoy will appear to have longer wave period than they do in a fixed frame of reference. This redshift, combined with negative spectral slope, $\partial E(f)/\partial f$, results in energy level that is reduced relative to what we would expect a buoy to measure from a fixed frame of reference.

The current study explores this hypothesis.

# 3. Methods

## *a. Frames of reference*

In this study, we consider three frames of reference:

1) The frame of reference that is fixed relative to the earth. We denote the frequency in this reference frame as $f_a$, being the absolute frequency. Since this is also our frame of reference for any velocity, the velocity of this frame of reference, $U_{fixed} \equiv 0$. In the literature, the associated circular frequency is often denoted as $\omega = 2\pi f_a$.
2) The frame of reference of the medium that the waves exist in. Following the conclusions of Pizzo et al. (2023) and Hara (2025), we treat this as the Lagrangian current—also known as the Lagrangian mean flow or Lagrangian drift—computed as the sum of the Eulerian current and Stokes drift[3]. We denote the frequency in this reference frame as $f_i$, being the intrinsic frequency or relative frequency. In the literature, the associated circular frequency is often denoted as $\sigma = 2\pi f_i$. This frequency is used to compute wavenumber $k$ using the linear dispersion relation, which in deep water is $\sigma^2 = gk$, where $g$ is gravitational acceleration. We assume that all waves in a spectrum are travelling in the same medium, though in reality, the vertical decay of their orbital motion is frequency-dependent[4]. This implies an assumption that there is no vertical shear in the Lagrangian drift. To draw attention to this treatment of the ocean as a moving slab (see Ortiz-Suslow et al. 2025), we denote the velocity of this medium as $U_{slab}$.
3) The frame of reference of a drifting buoy, $U_{drift}$. The buoy measures frequencies denoted $f_d$.

## *b. Methods of changing the frame of reference*

Two methods are applied in this study to transform the spectrum from the drifter frame of reference to the fixed frame of reference. In both cases, waves are assumed to travel in the wind direction. This is, in the mean, appropriate for high frequency wind sea. Further, the waves are assumed to have zero directional spread. In the real ocean, wind sea has significant directional spread (see, e.g., Donelan et al. 1985). In cases where the wind direction is aligned with the $\vec{U}$ used in the Doppler equation, this assumption implies that we will tend to over-estimate the Doppler effect. Possible improvements to this calculation are discussed in Section 5.

---

[3] This implies that the Stokes drift caused by the ocean waves contributes to the motion of the medium that the waves are in.

[4] In deep water, the orbital velocity decays as $e^{kz}$, where $z$ is the distance from the mean sea level, 'positive up'.

Both methods require the 10-m wind speed $U_{10}$ and/or vector $\overrightarrow{U_{10}}$. In the case of application to the Spotter buoys, co-locations of hourly, 1/4° ERA5 wind speed with the buoys are created using nearest neighbor in time and bilinear interpolation in space.

METHOD 1, USING WIND-BASED DRIFT

The first method is the simpler of the two applied here. In Step 1, we Doppler shift to intrinsic frequency using the Doppler equation,

$$2\pi f_2 = 2\pi f_1 + \vec{k} \cdot \overrightarrow{(U_1 - U_2)} \quad (1)$$

Here, $\vec{k}$ is wavenumber and subscripts denote two frames of reference. With the assumption that the waves, $U_{drift}$ and $U_{slab}$ are all in the wind direction, this simplifies to:

$$f_i = f_d + k(U_{drift} - U_{slab})(2\pi)^{-1} \quad (2)$$

or, in deep water,

$$f_i = f_d + (2\pi)f_i^2(U_{drift} - U_{slab})g^{-1} \quad (3)$$

The difference between $U_{drift}$ and $U_{slab}$ is assumed to be caused by windage on the buoy. The windage is approximated as 1% of $U_{10}$, following the finding of Houghton et al. (2021), for a Sofar Spotter buoy (see also Davis et al. 2025).

Equation (3) is a quadratic equation with two roots: e.g., see Drennan et al. (1994), Collins et al. (2017), and Colosi et al. (2023). The lower-valued root (closest to $f_d$) is selected. Wavenumber $k$ is thus solved. For our problem the spectrum $E(f_i)$ is not needed, so we proceed to the next step.

In Step 2, we Doppler shift to $f_a$. This can be done either from $f_d$ or from $f_i$. We show the latter approach here:

$$f_a = f_i + k(U_{slab} - U_{fixed})(2\pi)^{-1} \quad (4)$$

The spectral density $E(f_a)$ is then computed using the relation $E(f_a)df_a = E(f_d)df_d$ (e.g., Collins et al. 2017). These values of $E(f_a)$ are valid, but due to the Doppler shifting, the discretization of $f_a$ is inconvenient, since each record with a different relative velocity $U_r$ (being $U_{fixed} - U_{slab}$ in this case) will have a different discretization. For this reason, we remap $E(f_a)$ to a discretization of $f_a$ that matches that of $f_d$, using linear interpolation.

This method requires an estimate of the Lagrangian mean current, $U_{slab}$. For this, we use the "3% of wind rule", e.g., Samelson (2022), $U_{slab} = 0.03U_{10}$. With Step 1, this implies an assumption that $U_{drift} = 0.04U_{10}$.

Method 1 is demonstrated in Fig. 4 by applying it to a parametric spectrum for the case of $U_{10}$=10 m/s. This parametric spectrum was developed for high frequency waves and includes a transition from the equilibrium range spectral slope, $E(f) \propto f^{-4}$ , to a saturation range spectral slope, $E(f) \propto f^{-5}$ . It is described in Appendix A. Waves appear to have lower frequency in the $U_{drift}$ frame of reference compared to the $U_{fixed}$ frame of reference, so, in Fig. 4, the shift of $E(f_d)$ to $E(f_a)$ is a shift to higher frequencies, giving, for example, a factor 1.21 increase in $E(f)$ at 0.4 Hz. This factor $E(f_a)/E(f_d)$ at 0.4 Hz is plotted for a range of wind speeds in Fig. 5. Results for wind speeds below $U_{10}$=6 m/s are not available, since below this wind speed, the test frequency (0.4 Hz) falls outside the valid range of the parametric spectrum. In Fig. 5, two changes in slope are visible, associated with the $f^{-4}$ to $f^{-5}$ regime change in the two spectra, where they occur in $U_{10}$ space for 0.4 Hz.

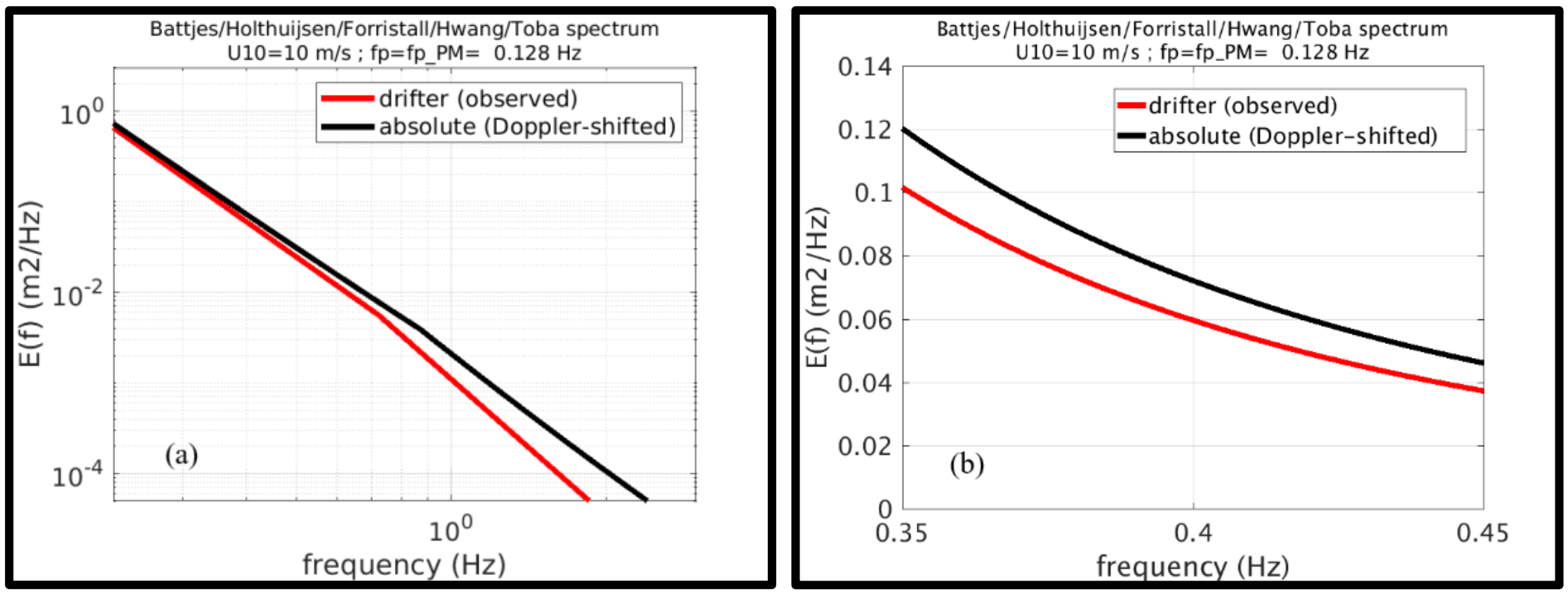

Fig. 4. Example of a parametric spectrum, Doppler shifted from a drifter frame of reference to a fixed frame of reference. The parametric spectrum is described in the text, and this example is for the case of $U_{10}$=10 m/s. Panel (a) shows this on log scale. Panel (b) shows the same spectra in the region of 0.4 Hz, on linear scale.

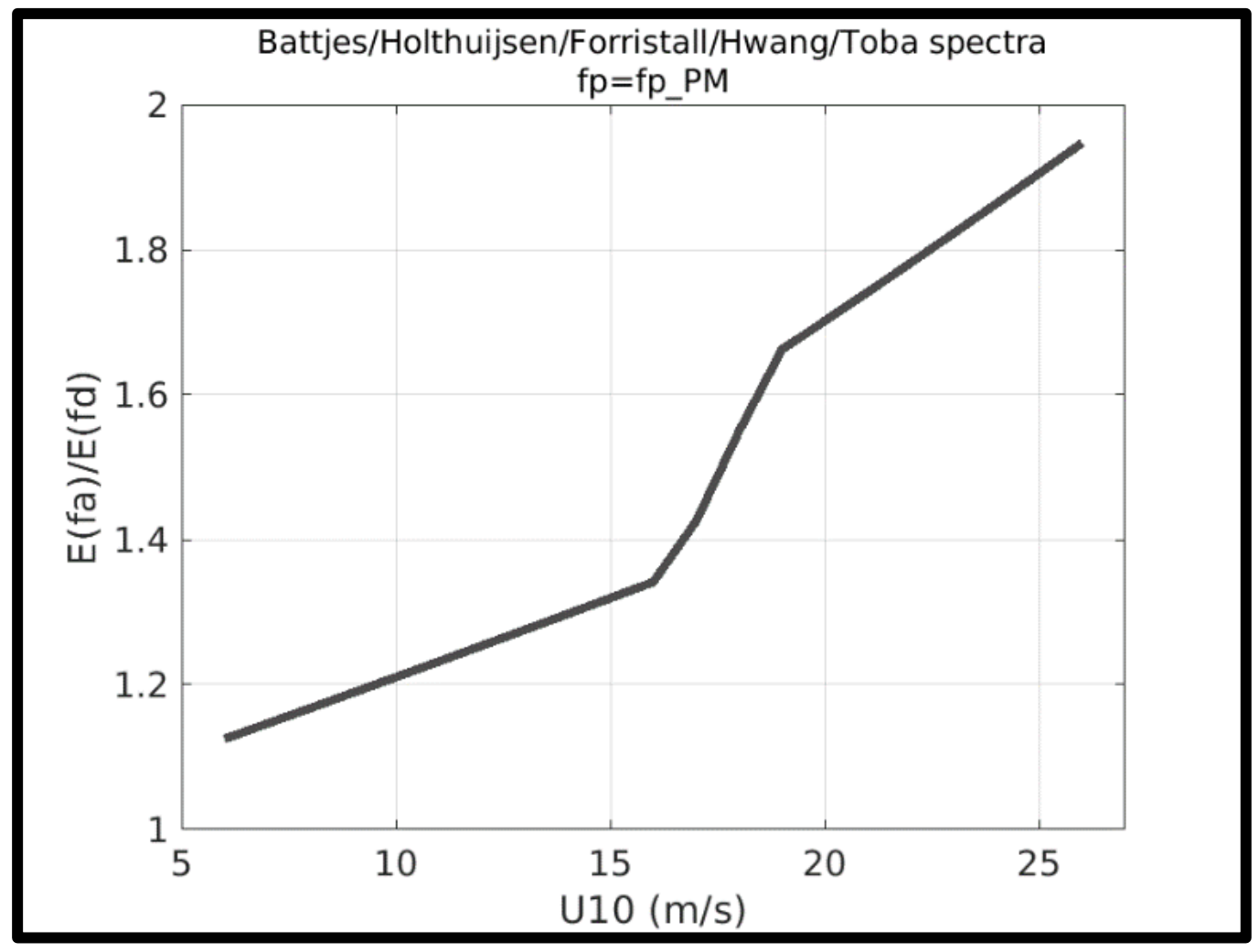

Fig. 5. The ratio $E(f_a)/E(f_d)$ vs. $U_{10}$ for the parametric spectra at 0.4 Hz.

METHOD 2, USING POSITION-BASED DRIFT

Step 1 of Method 2 is identical to that of Method 1. The relative velocity, $U_r = U_{drift} - U_{slab}$, is again assumed to be caused by windage only, and a simple function of wind speed, $U_r = 0.01U_{10}$.

In Step 2 of Method 2, we use an estimate of $U_{drift}$ based on buoy position, rather than using the empirical estimate $U_{drift} = 0.04U_{10}$ used in Method 1. The buoy positional change, $\overrightarrow{\Delta x}$, is computed, in meters, from the longitude and latitude of two buoy records. Centered differencing is used, e.g., $U_{drift,x} = (x(i+1) - x(i-1))/(t(i+1) - t(i-1))$, where $i$ denotes record number and $t$ is time in seconds. The buoy velocity is computed using the time elapsed between the two records, $\overrightarrow{U_{drift}} = \overrightarrow{\Delta x}/\Delta t$. Since we require the drift in the direction of the waves, and the wind direction is used as a proxy for the wave direction, this is converted to a scalar by projecting the vector along the axis of the wind, $U_{dr,sp} = (\overrightarrow{U_{10}} \cdot \overrightarrow{U_{drift}})/U_{10}$. This scalar can be negative, implying cases where the buoy is drifting against the wind (and waves). Quality-control procedures are used to eliminate records with cases of drift that may be spurious, e.g., if a buoy might be on a vessel or towed by a vessel. These are described in Appendix B.

In Method 2, we Doppler shift to $f_a$ from $f_d$:

$$f_a = f_d + k(U_{dr,sp} - U_{fixed})(2\pi)^{-1} \quad (5)$$

Fig. 6. compares the $U_{drift}$ used in Method 1 to the $U_{dr,sp}$ used in Method 2 as a two-dimensional histogram (i.e., scatter density plot). The number of co-locations is indicated on the plot, denoted $n$, and three error metrics, denoted CC, bias and RMSE. CC is the Pearson correlation coefficient. One obvious difference is that $U_{drift}$ is positive definite, while $U_{dr,sp}$ is sometimes negative. The empirical estimate $U_{drift}$ tends to be higher than $U_{dr,sp}$, with a bias of +0.11 m/s, and with the centroid of the scatter density plot being well above the 1:1 line. Based on this comparison, we can expect that application of Method 2 will result in smaller Doppler effect than application of Method 1.

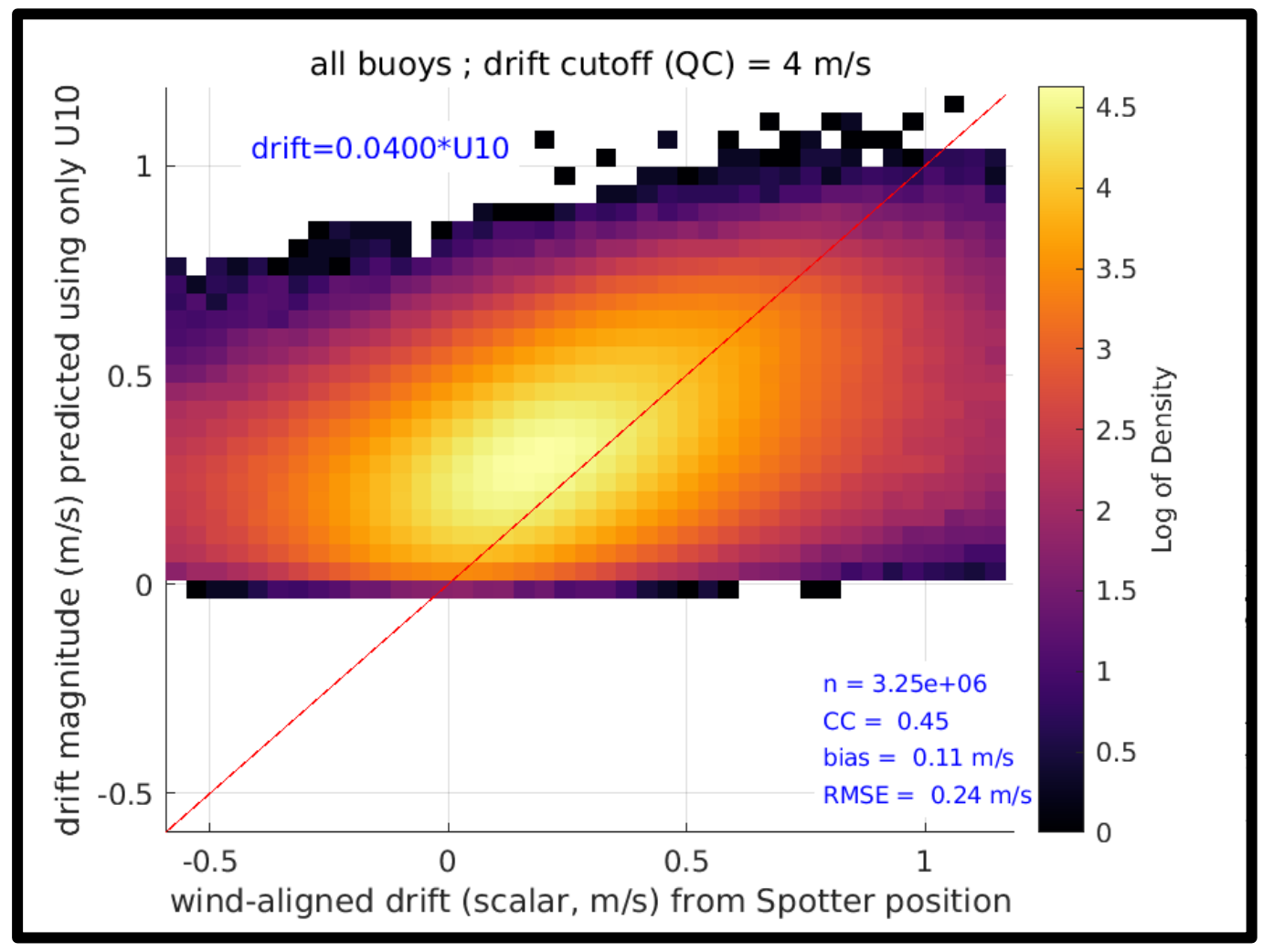

Fig. 6. A two-dimensional histogram comparing two ways to estimate drift of the Sofar buoys (9-month dataset). Vertical axis: drift estimated as 4% of $U_{10}$. Horizontal axis: drift estimated from rate of buoy position change along the axis of the wind. The thin red line is the 1:1 line.

## 4. Results

In this section, we apply the Doppler shift Methods 1 and 2 to the Sofar dataset and report the outcome. Each buoy spectrum is Doppler shifted individually and averaged[5]. In Fig. 7, the mean ratio $E(f_a)/E(f_d)$ vs. $U_{10}$ for the Sofar buoys are compared against the ratios from the parametric model of Fig. 5. The comparison clearly indicates a positive monotonic relationship between magnitude of the Doppler shift and wind speed. The two Spotter buoy types are kept separate in this analysis, and there is significant difference between the two types, particularly at the high wind speeds. This may be caused by the difference in frequency resolution (with buoy type 2 being more reliable in this regard), the difference in population size (where buoy type 1 has the advantage), or a combination of the two. As expected, application of Method 2 results in smaller Doppler effect than application of Method 1. Less expectedly, the Method 1 Doppler shift from the buoy spectra is much stronger than the same method applied to the parametric spectra. Since the magnitude of $E_a(f_a)/E_d(f_d)$ is largely determined by the spectral slope $\partial E_d(f_d)/f_d$ (along with $k$ and $U_r$, of course), this discrepancy is likely caused by a steeper spectral slope in the buoy data vs. the parametric model.

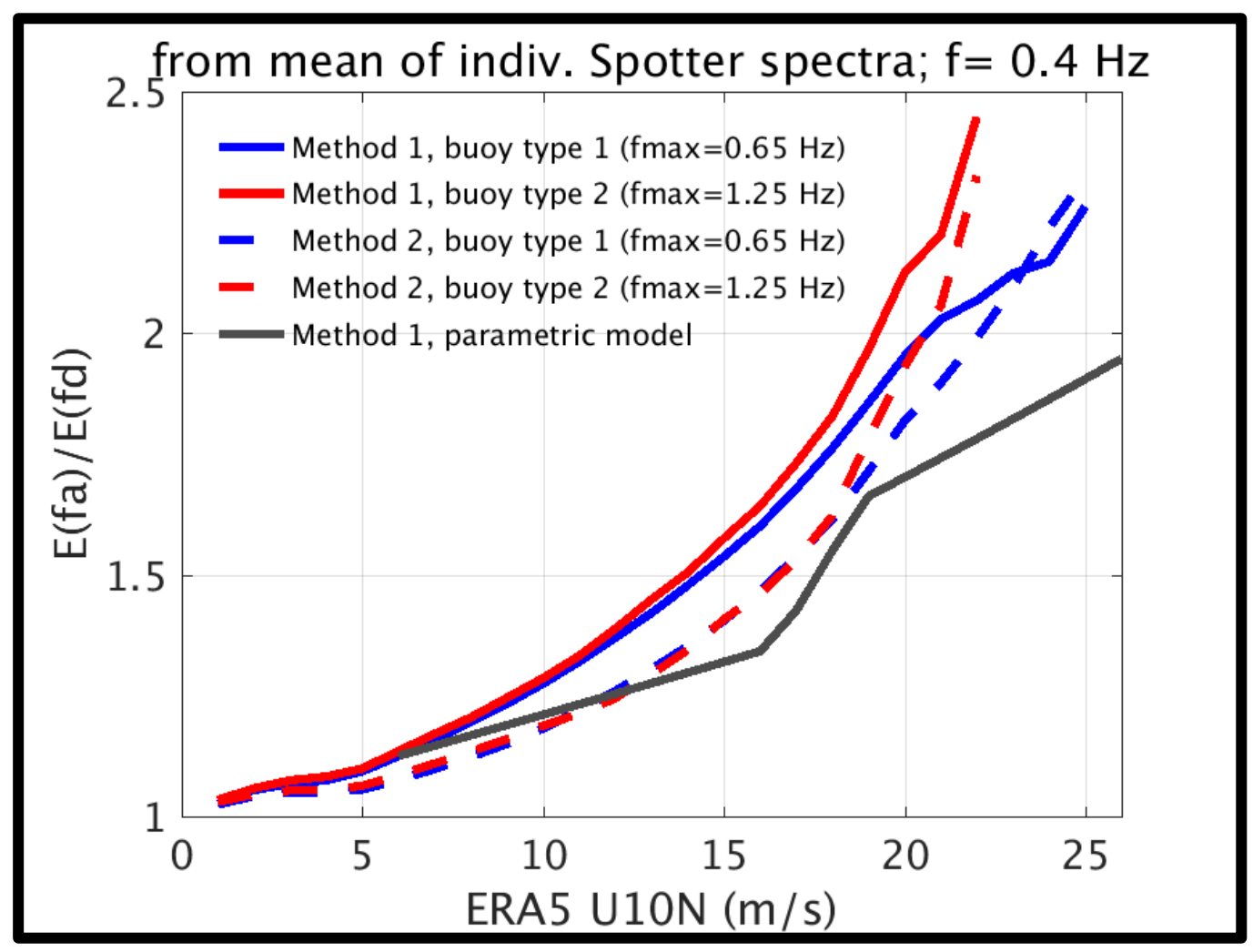

Fig. 7. The mean ratio $E(f_a)/E(f_d)$ vs. $U_{10}$ for the Sofar buoys (9-month dataset), using two methods (1 and 2) and two buoy types ($f_{max}=0.65$ Hz and $f_{max}=1.25$ Hz). The parametric model result from Fig. 5 is shown with a gray line.

In Fig. 8, we show the comparison from Fig. 3, $E(f)$ at $f$=0.4 Hz vs. $U_{10}$, with the Doppler-shifted Spotter values also plotted, for each of the two methods. The parametric spectrum is included for reference. After Doppler shifting, the Sofar spectra are much more consistent with the DWR spectra, especially using Method 1. It is reasonable to expect that the other drifting buoys (CORDC) would similarly affected, if Doppler shift was applied. By contrast, the NDBC buoys are already in a fixed frame of reference, so Doppler shift cannot explain the discrepancy between the NDBC buoys and DWR buoys: some other explanation is needed, such as the buoy response function. Similarly, since our calculations with the Sofar spectra indicate a significant Doppler shift at our test frequency, we can deduce that the good agreement between moored NDBC and the drifting buoys for $U_{10}$ =14 to 18 m/s is *not* a good sign. It is likely a fortuitous intersection of errors, where the effect of Doppler shift on the drifting buoys (in their drifting frame of reference, solid green line) has approximately the same effect as a—suspected, but not explored here—damped response of the NDBC buoys at high frequencies.

---

[5] In preliminary calculations, not shown here, we computed a "mean Spotter spectrum" for each wind speed and Doppler shifted that spectrum. The results are broadly similar to those presented here for $U_{10} < 20$ m/s but showed spurious behavior at higher wind speeds, so only the more careful approach is presented here.

As we showed in Fig. 7, the more painstaking method of estimating buoy drift (Method 2) results in less Doppler shift than Method 1. Thus, it can be reasonably argued that Method 1 represents a high estimate of the Doppler shift. Further, we assume that all high-frequency waves are wind-aligned, disregarding the reduction of Doppler shift by directional spread that is inevitable in the real ocean, which also causes our methods to yield a high estimate of the Doppler shift. With this being the case, we cannot take the excellent agreement between DWR buoys (black line) and Sofar Doppler shifted via Method 1 (dashed green line) to indicate that the problem is fully solved. Again, another explanation may be required to close the gap completely.

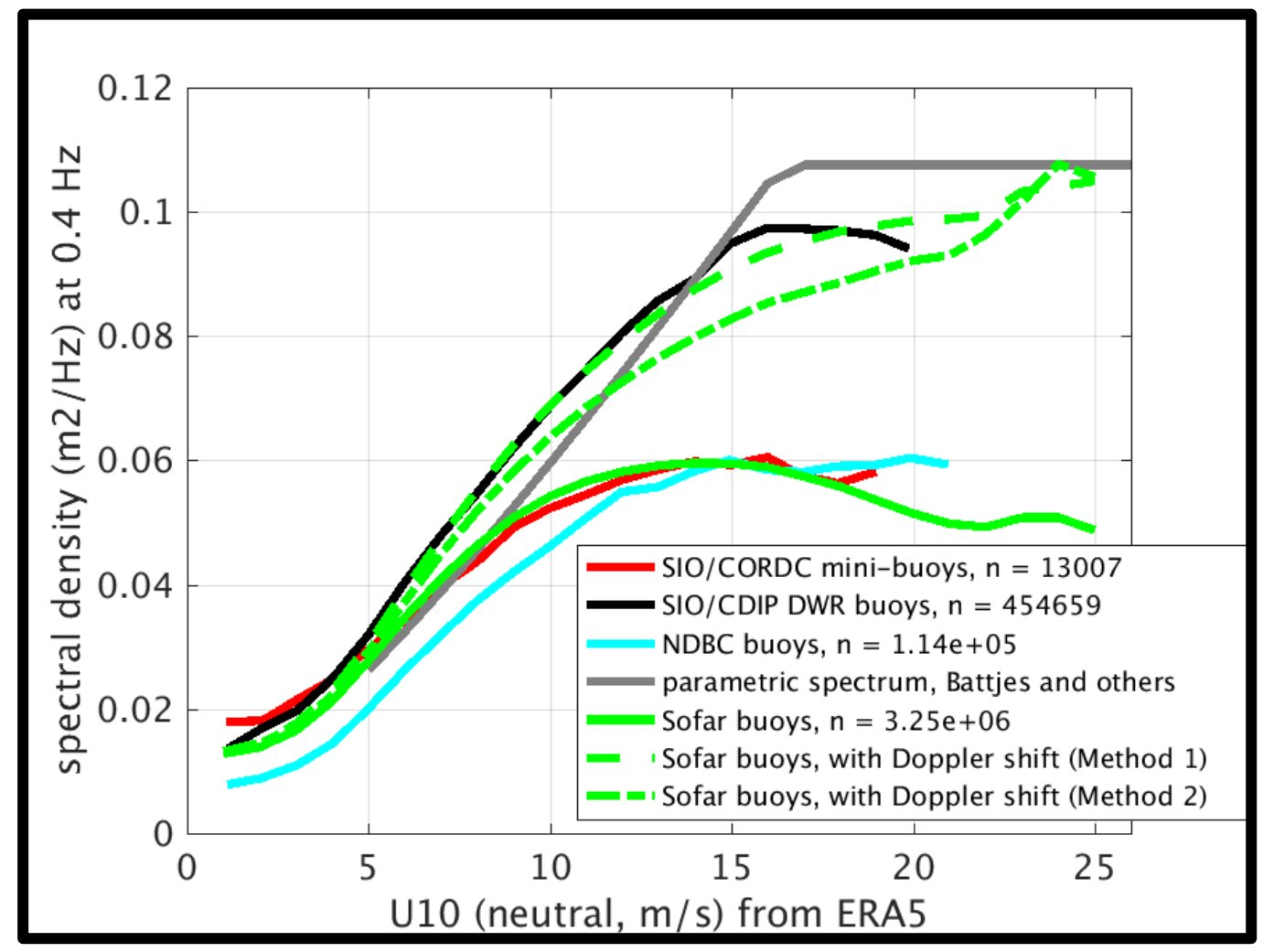

Fig. 8. $E(f)$ at $f$=0.4 Hz vs. $U_{10}$ for several buoy datasets. In the case of the Sofar dataset, Doppler-shifted results (fixed frame of reference) are shown, using two methods for estimating buoy drift (dashed lines), and original Spotter spectra (drifting frame of reference) are shown with a solid green line. Values from the parametric spectra without Doppler shift are also shown.

## 5. Discussion

### *a. Doppler shift using alternative position-based method*

In this paper, our Method 2 of Doppler shifting from $E_d(f_d)$ to $E_a(f_a)$ uses the position-based estimate of drift, $U_{drift}$ in its Step 2. A third method of Doppler shifting was tested that also uses $U_{drift}$, but applies it in Step 1, during the calculation of $f_i$. Thus, the position-based estimate of drift affects the wavenumber $k$ that is used in the Doppler equation in Step 2.

However, this is the only change to Step 2. The relative velocity $U_r = U_{drift} - U_{fixed}$ is the same as that applied in Method 2. We found only modest difference in outcome relative to Method 2, so results from this third method are not presented here.

*b. Doppler shift using wave directional information*

Our methods here assume that the wind and waves are aligned without directional spread. In the context of high-frequency waves, the latter assumption is of primary concern. For buoy observations, the normalized directional distribution at a given frequency, $D(\theta)$, is typically described using Fourier coefficients $a_1$, $b_1$, $a_2$, $b_2$. For definition, see Kuik et al. (1988), Appendix A. Kuik et al. (1988) recommended that "reconstruction of $D(\theta)$ from pitch-and-roll buoy data should therefore be undertaken only in a nonroutine matter, if the distribution of wave energy over directions is a strictly required quantity in any further processing". We agree with this as "best practices" but unfortunately, this is one of those situations where the application of a directional estimator is "strictly required": it is not possible to Doppler shift the buoy-derived parameters, $E(f)$, $a_1(f)$, $b_1(f)$, $a_2(f)$, $b_2(f)$ without the estimator, because the Doppler shift acts across $(f, \theta)$ space simultaneously. Application of a directional estimator for this purpose was, to our knowledge, first done by Forristall et al. (1978), using a relatively primitive estimator, applied to the case of Doppler shift of waves by currents. Drennan et al. (1994) make a similar calculation to correct for Doppler shift by ship motion. Colosi et al. (2023) apply an information-preserving directional estimator (Maximum Entropy Method, Lygre and Krogstad 1986) to wave measurements by ASVs. Hanson et al. (1987) create an advanced directional estimator and apply it to the ship problem, but their method requires a spatial array and thus is not applicable to the pitch-and-roll buoy problem.

*c. Implications for parametric spectra*

The results of this study strongly indicate that high-frequency spectra measured by drifting buoys must be Doppler shifted prior to applications to design of parametric spectra, particularly in medium-to-high wind conditions.

*d. Follow-up work*

A direct follow-up to the present study was indicated earlier in this Discussion. We suggest that the study is repeated, accounting for directional spread, applying the Doppler shift to the two-dimensional spectra.

We also recommend that the observational data, both new and historical, be remapped to a common frame of reference. The intrinsic frequency spectrum (thus, $U_{slab}$ frame of reference) is the most obvious choice, since it permits direct comparison to spectra measured by satellites. This involves a major challenge, since $U_{slab}$ must be defined. Any estimate, e.g., from a global circulation model, will have some level of uncertainty, and the accuracy of the Doppler-shifted spectrum will be tied to the accuracy of the $U_{slab}$ estimate.

## 6. Summary of conclusions

We find that high frequency energy levels from moored Datawell Waverider (DWR) buoys are significantly higher than those from drifting wave buoys (Fig. 2, 3). Doppler shifting is proposed as one of two hypotheses to explain the discrepancy. We introduce two methods for Doppler shifting wave spectra from a drifting frame of reference to a fixed frame of reference. This required a new procedure that accounts for three frames of reference (fixed, intrinsic, and drifting) rather than two. Both methods use a simple estimate for windage caused by wind drag on a buoy. They differ in that the first uses a simple wind-based estimate of Lagrangian drift (Fig. 4, 5), while the second computes total drift from buoy positions. Both methods assume that high-frequency waves are aligned with the wind, with zero directional spread. Thus, it is expected to be a high (upper limit) estimate of Doppler shift. The first method, entirely wind-based, yields a higher estimate of buoy drift than the second which uses positional information (Fig. 6). Thus, the first method results in larger effect of Doppler shift than the second.

Applying the two Doppler-shift methods to over three million spectra reported by Sofar Spotter drifting buoys, we find a large impact on high frequencies, particularly at higher wind speeds. For our test frequency of 0.4 Hz, there is a 50% increase of spectral density at $U_{10}$ =14-18 m/s and a 100% increase at $U_{10}$ =19-22 m/s (Fig. 7).

The Doppler shift increases the Sofar energy level at the test frequency to be in much better agreement with the moored DWR spectra (Fig. 8). Thus, Doppler shift appears to be the primary factor contributing to the discrepancy between the drifting buoys and DWR buoys. Given that our estimate of Doppler shift is high (via omission of directional spread), it is likely that secondary factors also contribute to the observed discrepancy.

We found that energy at our test frequency created using a numerical model hindcast is broadly consistent with the levels reported by the DWR buoys, and thus broadly inconsistent with the levels reported by the drifting buoys. We also found that high frequency energy levels from National Data Buoy Center (NDBC) buoys are significantly lower than those from DWR buoys (Fig. 3, 8). Since both are moored, Doppler shift is not a likely explanation for the difference. There is good correspondence in energy levels between NDBC and the drifting buoys for $U_{10}$=14 to 18 m/s, but taken in context of our other results, this agreement is probably spurious, i.e., a result of cancellation of errors.

*Acknowledgments.*

This work was funded by the Office of Naval Research through the NRL Core Program, Program Element 0602435N. It was funded by NRL Core 6.2 project, PI E. Rogers, Work Unit 62A1P9. This manuscript is approved for public release, STRN: NRL/7320/JA--2026/17.

We thank Pieter Smit and others at Sofar Oceans for the Sofar Spotter dataset. We are grateful for observational datasets used in this study that are provided freely to the research community, and acknowledge the significant effort and resources required to make this available. We thank Eric Terrill, Mark Otero, Peter Rogowski, and others at CORDC for the UCSD/SIO/CORDC mini-buoy dataset. We thank the engineers at APL/UW for the OSP wave buoy dataset. We thank James Behrens and others at CDIP for the UCSD/SIO/CDIP datasets. We thank NOAA/NDBC for data from their moored buoy network. We thank ECMWF and the European Commission for sharing their ERA5 wind and ice products through C3S of the Copernicus Earth Observation Programme. We thank Drs. David Wang (NRL), Jie Yu (NRL), Eric Terrill (CORDC), and Sophia Merrifield (CORDC) for valuable insights, suggestions, information, and discussion.

*Data Availability Statement.*

NOAA NDBC data were obtained from https://www.ndbc.noaa.gov/ and UCSD CDIP data were obtained from https://cdip.ucsd.edu/ . Both are publicly available. ERA5 winds were obtained from https://cds.climate.copernicus.eu/datasets/reanalysis-era5-single-levels. This is also available to users with accounts, which are free. The UCSD CORDC and Sofar datasets are proprietary, but the authors are happy to share the data if the requestor can verify permission from those originators.

APPENDIX

## Appendix A. Parametric spectrum

In this manuscript, a parametric spectrum is used for illustration and comparison. To be clear, the parametric spectrum does *not* have a role in the transformation of buoy spectra.

The equilibrium range of the spectrum (Resio et al. 2004) is based on Toba (1973):

$$E(\sigma) = \beta g u_* \sigma^{-4}$$

or

$$E(f) = \beta g u_* (2\pi)^{-3} f^{-4} \tag{6}$$

For the proportionality coefficient, we use $\beta$=0.096, following Holthuijsen (2007), who takes it as an average of field observations tabulated by Battjes et al. (1987). The friction velocity $u_*$ is computed using Hwang (2011):

$$u_* = U_{10}\sqrt{10^{-4}(-0.0160U_{10}^2 + 0.967U_{10} + 8.058)} \tag{7}$$

The parametric spectrum only defines the high frequency portion of the spectrum. It is treated as invalid for $f < 1.5f_p$, where $f_p$ is the peak frequency. Above a transition frequency $f_t$, the slope of the spectrum changes such that $E(f) \propto f^{-5}$, known as the saturation range, following Tsagareli (2009) and Vincent et al. (2019). In our case, this simplifies to:

$$E(f) = \beta g u_* (2\pi)^{-3} f_t f^{-5}$$

The peak frequency is $f_p = \psi f_{p,PM}$ where $f_{p,PM}$ is the "fully developed" value given by Pierson and Moskowitz (1964), $f_{p,PM} = 0.13g/U_{10}$. The setting $\psi$=1 is used herein, but is more generally $\psi \geq 1$.

The transition frequency $f_t$ is taken from Forristall (1981): $f_t = 0.29/u_*$.

## Appendix B. Quality controls on buoy data

Here we describe quality control (QC) methods applied to the Sofar buoy dataset.

*a. Band height: grounded buoys*

The $H_{m0B}$ values computed from Sofar spectra, shown in Fig. 1, were filtered to remove cases where it was particularly obvious that the buoy was grounded or within sea ice. These were records for which $U_{10} > 2.5$ m/s and $H_{m0B} < 0.03$ m.

*b. Spectral density: grounded buoys*

The $E(f)$ data at test frequency $f = 0.4$ Hz were not filtered specifically to remove cases of obviously grounded buoys, since other filtering made it unnecessary. We verified this by looking for cases with $U_{10} > 4.3$ m/s and $E(f) < 1 \times 10^{-4}$ $m^2$/Hz; we found none.

*c. Drift speed computed from buoy positions*

The following criteria were applied:

1) A cut-off of $E(f) < 1 \times 10^{-5}$ m$^2$/Hz is applied. Where the spectral density is below the cut-off, the drift speed is considered invalid for that record, and also for the prior and following record. This is intended to catch cases where the buoy is on a ship rather than in the water.
2) The original data files are organized by month. In cases where the time series begins or ends on a date not near the beginning or end of the month, the drift speed is considered invalid for the first (or last) 12 hours of the time series. This is intended to find cases where the buoy is on a ship rather than in the water. This will not catch cases for which the deployment/retrieval is near the beginning or end of the month.
3) Records with high computed drift ($U_{drift} > 3.5$ m/s), drift speed are considered invalid. The QC is also applied to the two records prior and following to the record with high drift (thus, up to 5 records are flagged). This is intended to find cases where the buoy is either on or pulled behind a ship and cases where the buoy is caught in unusually strong coastal currents.

*d. QC applied during Doppler-shift calculations*

For both methods and all cases, if the spectral density $E(f_d)$ or $E(f_a)$ at the test frequency $f = 0.4$ Hz is very low (less than $1 \times 10^{-10}$ m$^2$/Hz), the record is omitted.

For the wind-based drift method, a cutoff of $1 \times 10^5$ is applied to the ratio $E(f_a)/E(f_d)$. In cases of larger ratios, the case is discarded.

For the position-based drift method, cases with spectral density $E(f_d)$ less than $1 \times 10^{-3}$ m$^2$/Hz are omitted. Cases with invalid drift estimates are, of course, also omitted.

Because of the differences in QC for the two methods, the populations of the two applications are not identical: For the wind-based drift method, $n = 3.250 \times 10^6$ and for the position-based drift method, $n = 3.243 \times 10^6$.